\documentclass[12pt,preprint]{aastex}
\usepackage{epsfig}
\newcommand{\impc}{\hbox {\rm Mpc}^{-1} }
\def\himsun{{h^{-1}M_\odot}}

\def\kms{\,{\rm {km\, s^{-1}}}}

\def\mpc{\,h^{-1}{\rm {Mpc}}}
\def\kpc{\,h^{-1}{\rm {kpc}}}

\shorttitle{Dark Halo Structure}
\shortauthors{Zhao et al.}
\begin{document}
%
%%%%%%%%%%%%%%%%%%%%%%%%%%%%%%%%%%%%%%%%%%%%%%%%%%%%%%%%%%%%%%%%%%%%%%
\title{Mass and redshift dependence of dark halo structure}

%%%%%%%%%%%%%%%%%%%%%%%%%%%%%%%%%%%%%%%%%%%%%%%%%%%%%%%%%%%%%%%%%%%%%%
\author{D.H. Zhao${^{1,2}}$, Y.P. Jing${^{1,2}}$,
H.J. Mo${^{3}}$, G. B\"orner${^{2}}$}
\altaffiltext{1}{Shanghai Astronomical Observatory, the Partner Group of
MPI f\"ur Astrophysik, Nandan Road 80,  Shanghai 200030, China}
\altaffiltext{2}{Max-Planck-Institut f\"ur Astrophysik,
Karl-Schwarzschild-Strasse 1, 85748 Garching, Germany; e-mail:
dhzhao@center.shao.ac.cn}
\altaffiltext{3}{Department of Astronomy, University of Massachusetts,
Amherst MA 01003, USA}

%%%%%%%%%%%%%%%%%%%%%%%%%%%%%%%%%%%%%%%%%%%%%%%%%%%%%%%%%%%%%%%%%%%%%%
%
%
\begin{abstract}
Using a combination of N-body simulations with different
resolutions, we study in detail how the concentrations of cold dark
matter (CDM) halos depend on halo mass at different redshifts. We
confirm that halo concentrations at the present time depend strongly
on halo mass, but our results also show marked differences from the
predictions of some early empirical models. Our main result is that
the mass dependence of the concentrations becomes weaker at higher
redshifts, and at $z\ga 3$ halos of mass greater than
$10^{11}\himsun$ all have a similar median concentration, $c\sim
3.5$. While the median concentrations of low-mass halos grow
significantly with time, those of massive halos change only little
with redshifts. These results are quantitatively in good agreement
with the empirical model proposed by Zhao et al.  which shows that
halos in the early fast accretion phase all have similar
concentrations.
\end{abstract}

%%%%%%%%%%%%%%%%%%%%%%%%%%%%%%%%%%%%%%%%%%%%%%%%%%%%%%%%%%%
\section{Introduction}
%%%%%%%%%%%%%%%%%%%%%%%%%%%%%%%%%%%%%%%%%%%%%%%%%%%%%%%%%%%

High-resolution $N$-body simulations have shown that the density
profiles of cold dark matter (CDM) halos can be described
reasonably well by a universal form,
\begin{equation}
\label{eq:nfw}
\rho(r) = \frac{4\rho_s}{(r/r_s)\left(1+r/r_s\right)^2}\,,
\end{equation}
where $r_s$ is a characteristic ``inner'' radius, and $\rho_s$ is the
density at $r_s$ (Navarro, Frenk and White 1996, 1997; NFW hereafter),
although there is still debate about the exact value of the
inner slope (e.g. Fukushige \& Makino 1997, 2003; Moore et al. 1998;
Jing \& Suto 2000).
For a halo of radius $R$, this profile is characterized by the
concentration parameter $c=R/r_s$, which is found to be dependent on
halo mass (smaller halos have, on average, higher concentrations). NFW
developed a simple model to account for this mass dependence. The NFW
model for the mass dependence of $c$ was later found to be
inconsistent with results obtained from a simulation of the
concordance low-density CDM model (LCDM) at high redshift
(Bullock et al., B01 hereafter). B01 found that $c$ is proportional to
the cosmic scale factor $a$ for a given halo mass. Based on this, B01
and Eke et al. (2001; E01 hereafter) proposed new empirical
prescriptions to predict $c$ as a function of redshift and halo mass.

It must be noticed, however, that all these prescriptions only give the
mean concentration of all halos of a given mass at a given
redshift. Since the density profiles are found to vary significantly
from one halo to another even for a given halo mass (Jing 2000; B01),
it is important to have a recipe to predict $c$ for individual halos.
Jing has examined the density profiles for halos in different dynamical
states, and found that the halo density profiles are closely
related to the halo formation history. The connection between halo
concentration and halo formation history was explored further by
Wechsler et al. (2002; W02). Assuming that $c=4.1$ at their defined
``formation redshift'' and $c\propto a$, W02 found that their model
prediction for $c$ is in good agreement with their simulation results
for the LCDM model.

In a recent paper, Zhao et al. (2003; hereafter ZMJB) found that, for
a given halo, there is a tight correlation between the inner scale
radius $r_s$ and the mass within it, $M_s$, for all its main
progenitors, and that this correlation can be used to predict the
concentration of a dark halo at any time without making any {\it ad
hoc} assumption about the form of the mass accretion history. The ZMJB
model predicts that the evolution of $c$ of individual halos are not just
a function of $a$ (such as $c\propto a$ as W02 assumed), but tightly
connected to their mass growth rate: the faster the mass grow, the
slower the $c$ increase.

In this {\it Letter} we use a combination of high-resolution
simulations of different boxsizes to directly explore the halo
structures for a wide range of halo mass in a wide range of
redshifts. This allows us to study in detail how halo concentration
depends on halo mass at various redshifts, and to test the accuracy of
the various empirical models mentioned above. We will show that at
high redshift the simulated mass dependence of $c$ is much different
from some previous results, and is quantitatively in good agreement
with ZMJB prediction.

%%%%%%%%%%%%%%%%%%%%%%%%%%%%%%%%%%%%%%%%%%%%%%%%%%%%%%%%%%%%%%%%%%%%%%%
\section{Simulation results}
%%%%%%%%%%%%%%%%%%%%%%%%%%%%%%%%%%%%%%%%%%%%%%%%%%%%%%%%%%%%%%%%%%%%%%%

The cosmological simulations used in this paper are generated with a
parallel-vectorized Particle-Particle/Particle-Mesh code (see Jing \&
Suto 2002). The concordance CDM model with the
density parameter $\Omega_0=0.3$ and the cosmological constant
$\lambda_0=0.7$ is considered. The linear power spectrum has the shape
parameter $\Gamma=\Omega_0 h=0.20$ and the amplitude $\sigma_8=0.9 $,
where $h$ is the Hubble constant in $100 \kms \impc$, and $\sigma_8$
is the rms top-hat density fluctuation within a sphere of radius
$8\mpc$ at the present time. We use $256^3$ particles for the
simulation of boxsize $L=25\mpc$, and $512^3$ particles for the other
two simulations of $L=100$ and $300\mpc$ (Table 1). The simulations
with $L=25 \mpc$ and $100\mpc$ have been evolved by 5000 time steps
with a force softening length $\eta$ (the diameter of the S2
shaped particles, Hockney \& Eastwood 1981) equal to
$2.5 h^{-1}{\rm kpc}$ and $10 h^{-1}{\rm kpc}$, respectively.
As a result, there are
many halos with more than $3000$ particles in these simulations, and
these halos are resolved similarly to or better than the individual
halo simulations in early studies of the density profiles
(e.g. NFW). It has been shown that the resolution at this level is
sufficient for determining the concentration parameter (Jing 2000;
E01), though it may not be good enough for addressing the issues with
regard to the slope of the density profile in the central region of a
halo (e.g. Fukushige \& Makino 1997, 2003; Moore et al. 1998; Jing \& Suto
2000). The simulation of $L=300\mpc$ is a typical cosmological
simulation, evolved by 1200 steps and with a force softening length
of $30 h^{-1}{\rm kpc}$. The halo sample constructed from
these simulations is big, which is essential for accurately
determining the mean halo concentration. There is a sufficient overlap
in mass between the halos of different simulations, from which the
resolution effect can be reliably estimated. The halos are defined
according to the spherical virialization criterion (Kitayama \& Suto
1996; Bryan \& Norman 1998), so the radius $R$ of a halo in this paper
is the virial radius $r_{\rm vir}$. The halos are identified from
simulations using the potential minimum method as described in Jing \&
Suto (2002), and the particle with the minimum potential in each halo
is chosen as the halo center. We use all halos identified
this way without applying any further selection criteria.

\begin{table}
  \centering
  \caption{A summary of simulation parameters\label{tab:sim}}
\begin{tabular}{lcccccc}
\tableline
  simulation & $N_{\rm p}$ & box size & $M_{\rm part}$ & $\eta$ &$z_
  {\rm initial}$\\
    & & $\mpc$ & $\himsun$ & $\kpc$ & & \\ \hline
  LCDM025  & $256^3$  & 25  & $8.0\times10^{7}$  & 2.5&72\\
  LCDM100  & $512^3$  & 100 & $6.4\times10^{8}$  & 10 &72\\
  LCDM300  & $512^3$  &300  & $1.7\times10^{10}$ & 30 &36 \\
\hline
  \end{tabular}
\end{table}

%%%%%%%%%%%%%%%%%%%%%%%%%%%%%%%%%%%%%%%%%%%%%%%%%%%%%%%%%%%
\subsection{The mass accretion history}

Following ZMJB, we construct the main branch of the merger tree for
each halo identified at redshift $z=0$, and work out the mass
accretion history. In Figure 1, we plot the mass accretion
histories for 20 randomly selected halos at each of the following mass
scales: $M_{\rm vir,0}=7\times 10^{10}\himsun$, $1.3\times 10^{13}\himsun$, and
$1.4\times 10^{15}\himsun$. As in ZMJB, the mass accretion history of
each halo is divided into a fast accretion phase and a slow accretion
phase, and we denote the transition redshift between the two by
$z_{tp}$.  With the mass $M_{\rm vir}(z)$ in units of $M_{\rm vir,tp}
=M_{\rm vir}(z_{\rm tp})$
and the physical virial density $\rho_{\rm vir}(z)$ in units of
$\rho_{\rm vir,tp}=\rho_{\rm vir}(z_{\rm tp})$
\footnote{For the LCDM model, $\rho_{\rm vir}(z)$ is 180 times at high
redshift and 101 times at $z=0$ of the critical density.  Both
$\rho_{\rm vir}(z)$ and $1+z$ can be used to denote the cosmic time
for a given cosmology.  We found the relation between $M_{\rm vir}$ and
$\rho_{\rm vir}(z)$ is better behaved than the $M_{\rm vir}$ - $z$ relation,
especially for low-density universes at $z\sim 0$.}, we found that the
mass accretion history has a universal form for halos of different
mass, and the mean accretion history can be accurately represented by
the thick smooth line for the halo masses covered by the simulations
(from $7\times 10^{10}\himsun$ to $1.3\times 10^{15}\himsun$).
Mathematically, the average accretion history can be expressed in the
form:
\begin{equation}
{M_{\rm vir}(z) \over M_{\rm vir,tp}}={x^{0.3}\over 1-a+ax^{-1.8a}}\,,
\label{mah}
\end{equation}
where $x=\rho_{\rm vir,tp}/\rho_{\rm vir}(z)$ and $a=0.75$
($0.42$) for the fast (slow) accretion phase. The universal
mass history has interesting implications for galaxy formation and
halo structure formation. Here because of the limited space, we
will only discuss its implications for halo structure in \S 3.

\begin{figure}
\epsscale{1.0} \plotone{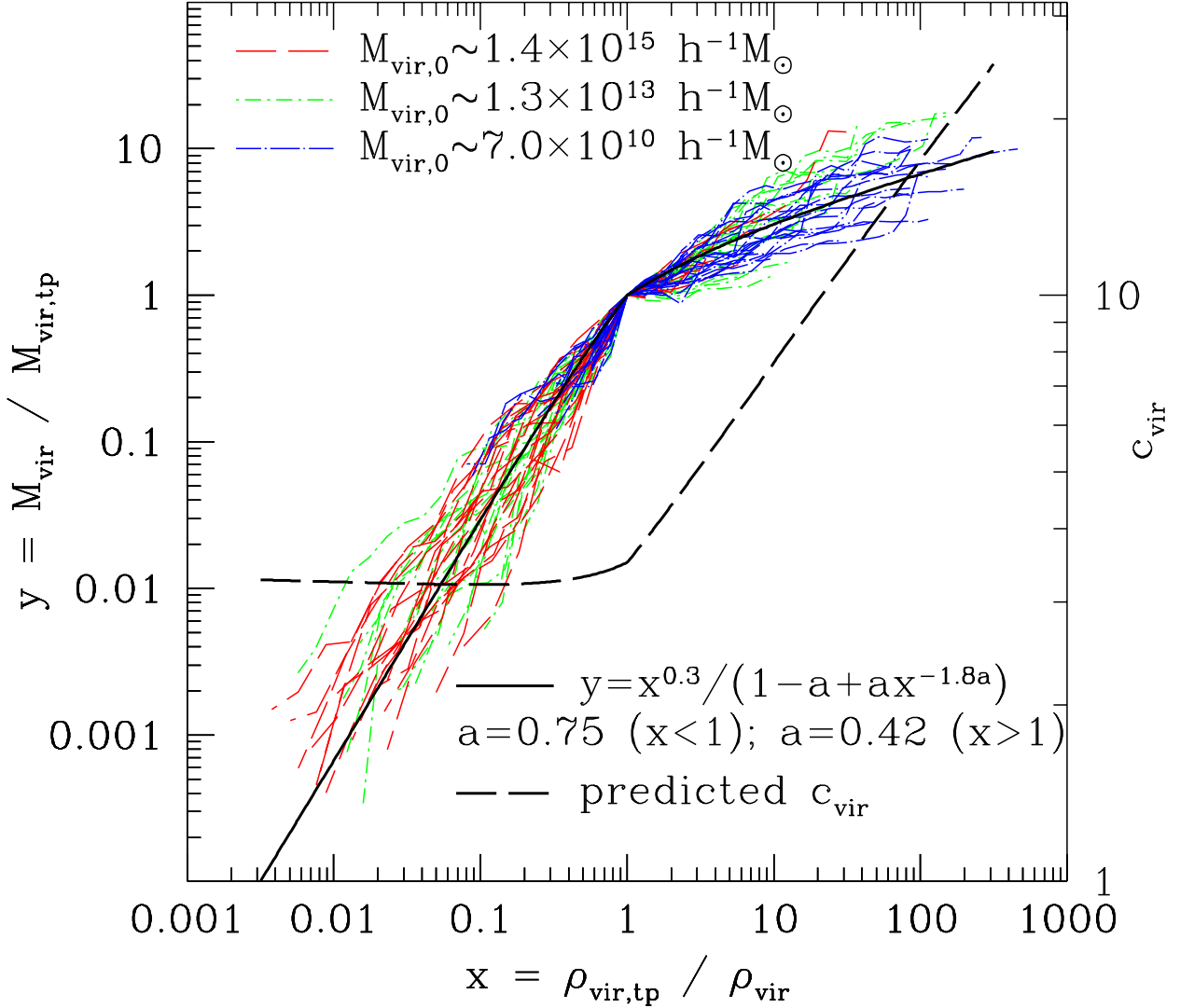} \caption{The mass accretion history
of dark matter halos along the main branch. Twenty halos are
randomly selected from the simulations at each mass indicated at the
top. We use the physical density within the virialized halo
$\rho_{\rm vir}(z)$ as the time variable. Both the halo mass $M_{\rm
vir}(z)$ and $\rho_{\rm vir}(z)$ are scaled to their quantities at
the turning point, $M_{\rm vir, tp}$ and $\rho_{\rm vir, tp}$. The
scaled mass accretion history on average is well represented by
Eq.(2) independent of the halo mass. The thick smooth dashed line is
the prediction of ZMJB for the concentration $c_{\rm vir}$ according
to the mean accretion history (the right vertical coordinate for
$c_{\rm vir}$). \label{fig:mah}}
\end{figure}

\subsection{Halo concentrations}

We select halos with more than $800$ particles at redshifts $z=0$,
$1$, $2$ and $4$, and determine the concentration parameter for each
of them by fitting the density profile to the NFW form (see ZMJB for
the fitting procedure). The halos are grouped in mass bins of
$\Delta \log_{10} M_{\rm vir} = 0.2$, and the median concentration
is calculated for each mass bin. The median concentrations
determined this way are presented in Figure 2 for masses larger than
$6.4\times 10^{10} \himsun$, together with their errorbars (standard
deviation among different halos in the bin divided by the square
root of the halo number). Note that there is always quite a large
overlap in halo mass between simulations of different boxsizes, and
that the median concentrations from different simulations are in
good agreement in the overlapping mass range. This agreement
suggests that our results are not significantly affected by the
finite numerical resolution, because for a given halo mass both the
mass and force resolutions decrease with the increase of the
simulation boxsize.

\begin{figure}
\epsscale{1.0} \plotone{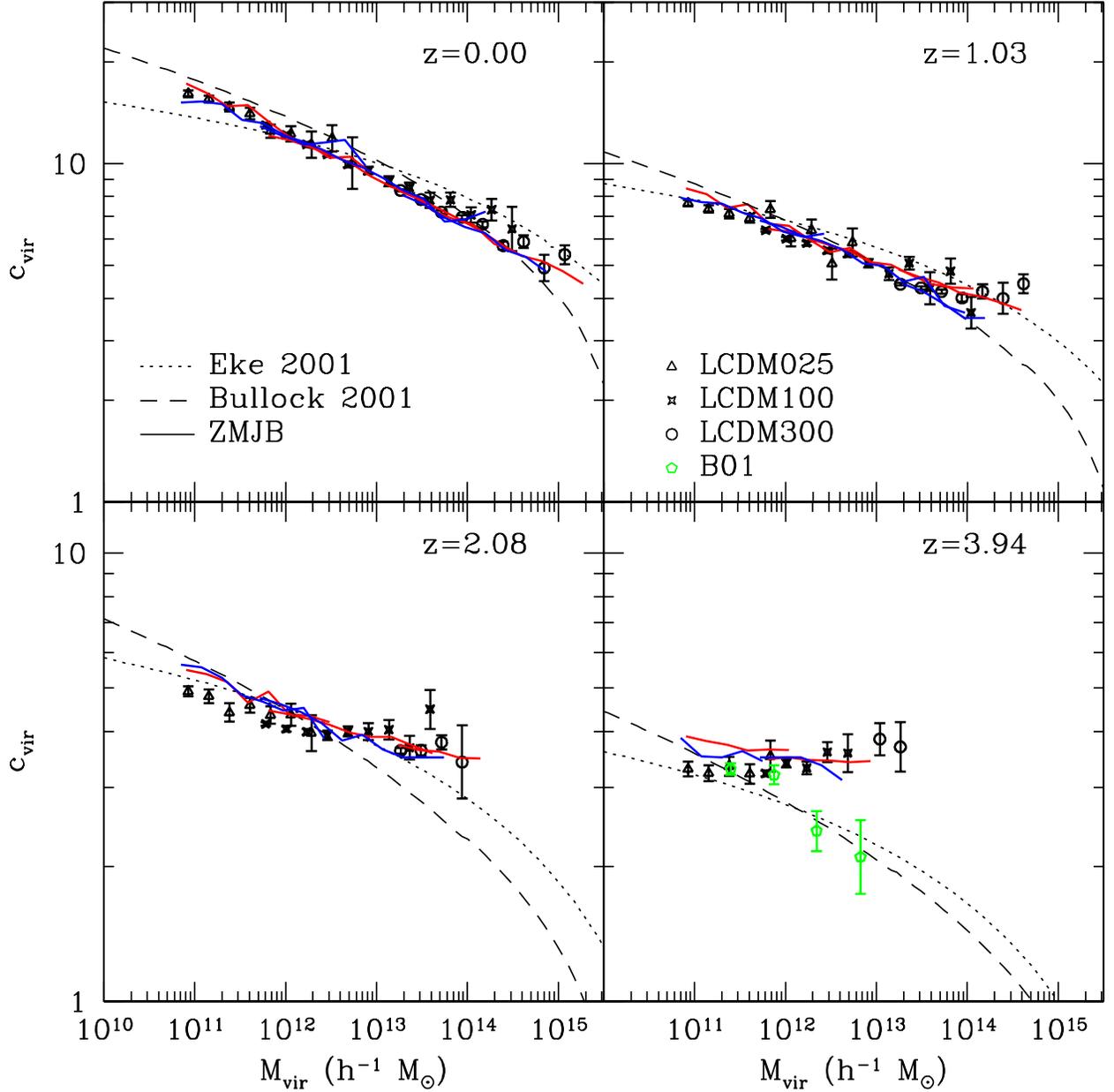} \caption{The median concentration of
halos as a function of the halo mass. Errorbars are the standard
deviation among halos of the same mass devided by square root of the
halo number. The black symbols are for the results measured in the
simulations, the blue and red lines are the predictions of ZMJB
using the mass accretion histories from the simulations (blue) and
from the PINOCCHIO model (red), and the dashed and dotted lines are
the predictions of the models of Bullock et al. and Eke et al.,
respectively. For comparison, the simulation results of Bullock et
al. for $z=4$ are plotted in the lower right panel as green
pentagons. \label{fig:cm}}
\end{figure}

As one can see, at low redshift, the median concentration decreases
rapidly with halo mass, from $c\sim 20$ for $M_{\rm vir}\sim 10^{11}\himsun$ to
$c\sim 4$ for $M_{\rm vir}\sim 10^{15}\himsun$.  This mass dependence is similar
to that found in earlier analyses (e.g. NFW; Jing \& Suto 2000; B01 ).
The mass dependence is weaker for halos at higher redshifts and
becomes insignificant at $z\ga 3$. This change in behavior is mainly
due to the fact that the median concentrations of small halos decrease
rapidly with increasing redshift while the concentrations of massive
halos change little.  Note that the decrease of $c$ with $z$ is slower
at higher $z$ and there seems to be a minimum value $c\sim 3$ for
the median concentration of dark halos.
This is true even for
a few halos at $z \sim 9$ that are not included here.

\section{Comparison with empirical models}

Based on results from numerical simulations, ZMJB found that the scale
radius $r_s$ of a halo and its scale mass $M_s$ (i.e. the mass within
$r_s$) are tightly correlated, with a relation well represented by a
simple power law:
\begin{equation}
\label{eq:msms0}
   {M_s \over M_{s,0}}=\left({r_s\over r_{s,0}}\right)^{3\alpha}\,,
\end{equation}
where $M_{s,0}$ and $r_{s,0}$ are the scale mass and scale radius at
some chosen epoch.
The value of $\alpha$ is found to be $0.52$ in the slow accretion
phase, and $0.64$ for the rapid accretion phase.  As
shown in ZMJB, this $M_s$-$r_s$ relation can be used to derive $c$
from the halo mass accretion history according to
\begin{equation}
\label{eq:cmh}
   {[\ln(1+c)-c/(1+c)]c^{-3\alpha} \over
   [\ln(1+c_0)-c_0/(1+c_0)]c_0^{-3\alpha}}=\left[{\rho_{\rm vir}(z)\over
   \rho_{\rm vir,0}}\right]^\alpha\left[{M_{\rm vir}(z)\over
   M_{\rm vir,0}}\right]^{1-\alpha}\,.
\end{equation}
This relation can be calibrated by fixing $c_{\rm tp}$ at
$z=z_{tp}$. ZMBJ have calibrated $c_{\rm tp}$ with five high resolution
halos, and adopted $c_{\rm tp}=4.0$. With our current large sample, we
find $c_{\rm tp}=3.5$ to be more accurate.

With the universal mass accretion history obtained from our
simulations, this recipe can be used to predict the median concentration
as a function of redshift. The result is shown as the smooth dashed
line in Figure 1. As one can see, the halo concentration has a value
about $3.5$ in the fast accretion phase, and scales roughly as $c\propto
\rho_{\rm vir}^{-1/3}(z)$ in the slow-accretion phase. This is
consistent with our simulation results, that halos with masses between
$10^{13}$ -- $10^{14}\himsun$ have median concentrations independent of
$z$ at $z\ga 1$. Most of these massive halos are in the fast
accretion phase at these high redshifts. Note that the median
concentration obtained from the simulations never drops below 3 even
for the most massive halos at the highest redshift probed by our
simulations, in agreement with the ZMJB model. The strong increase of
$c$ with decreasing $z$ for low-mass halos at low redshift observed in
the simulations is also consistent with the model prediction, because
most of those halos are in their slow accretion phases.

We generate samples of mass accretion histories from the simulations,
and apply the ZMJB model to predict the concentrations for each of
these halos (Figure 2). The model prediction reproduces well the mass
and redshift dependence of halo concentrations obtained from the
simulations. The distribution of the concentrations for given halo
mass and redshift is well described by the log-normal distribution
with $\sigma_{\ln c}\approx 0.3$ (Jing 2000, B01). We also compute
mass accretion histories using the PINOCCHIO code of Monaco et al. (2002).
This code identifies dark matter halos and their merging histories
by applying an ellipsoidal collapse model to an initial cosmic density
field. It has been shown to be quite accurate in reproducing many
properties of the halo population. All parameters are kept the same
as in the simulaitons, and also we apply the ZMJB model to predict the
halo concentrations. Again the agreement with our simulation
results is satisfactory with an accuracy better than $10 \%$
(Figure 2). In a forthcoming paper (Zhao et al., in preparation),
we will show that the ZMJB prediction is valid for a wide range of
cosmological models, including the SCDM model, an OCDM model, and
scale-free models.

The increase of the halo concentration with decreasing redshift in the
slow accretion phase is qualitatively consistent with the relation
$c(M_{\rm vir},z)\propto (1+z)^{-1}$ found by B01,
because $\rho_{\rm vir}(z)$ is approximately proportional
to $(1+z)^3$.  It is however important to note that the evolution
we obtained is along the main branches of merger trees, while the
relation $c(M_{\rm vir},z)\propto (1+z)^{-1}$ obtained by B01 is for
the median concentration of a halo population of a given mass
$M_{\rm vir}$. There is a marked difference of the prediction of the
ZMJB model from that of B01: while ZMJB predicts that $c$ does not
change in the fast accretion, B01 predicts $c\propto (1+z)^{-1}$
for a given mass.

In Figure 2, we compare our results to the predictions of the
empirical models given by B01 and E01.  The B01 model agrees with our
simulation results well at redshift $z=0$ for $M_{\rm vir}\le
10^{14}\himsun$. Note that this is approximately the mass range that
their simulation can effectively explore; their simulation uses
$256^3$ particles in a box of $60\mpc$. The B01 model also agrees with
the redshift dependence of $ c$ for low-mass halos, but it fails to
match our simulation results at the high mass ends. The model of E01
fits our simulation data better for $z=1$ and $2$, but worse for $z=0$
than the B01 model. Since the Eke et al. model adopted a
redshift-dependence of $c$ similar to that of B01 model, it also
underestimates the concentration for massive halos at high redshift.

 It should be pointed out that the models of B01 and E01
both match their own simulations to redshift $4$ and $2$,
respectively, and so there seems to be a discrepancy among
the different simulation results. Our results are
consistent with the simlation results of E01 over the mass and
redshift ranges probed by their simulations. Note that
the very low concentrations obtained by E01 are for a warm dark
matter spectrum. There is a marked difference between B01's
simulation results and ours for $z>2$; as comparison
we plot in the low-right panel of Figure 2 their simulation
results for $z=4$ (green pentagons). As one can see, the discrepancy
between B01 and our simulation results becomes significant
at $M_{\rm vir} \ga 5\times 10^{12}\himsun$ for this redshift.
Unfortunately, the origin of this discrepancy is unknown.
Our $100h^{-1}{\rm Mpc}$ box simulation has a mass resolution
slightly better than their main ($60h^{-1}{\rm Mpc}$) simulation,
and in terms of halo number, our sample is more than 4 times
larger. Since the number density of massive halos
is low at high $z$, and since halo concentration can
differ substantially for halos with the same mass at the
same redshift, a large sample might be crucial to get
reliable results. We are confident about our results, because
our halo sample is large and our simulations with different
resolutions and boxsizes agree with each other very well.

%%%%%%%%%%%%%%%%%%%%%%%%%%%%%%%%%%%%%%%%%%%%%%%%%%%%%%%%%%%%%%%%%%%%%%%%
%% If space is enough, could you please uncomment the sentence below? %%
%%%%%%%%%%%%%%%%%%%%%%%%%%%%%%%%%%%%%%%%%%%%%%%%%%%%%%%%%%%%%%%%%%%%%%%%
As mentioned earlier, the consistency of the results obtained for
different simulation boxsizes indicates that our determination of
halo concentrations should not be affected significantly by the
limited simulation resolution. As shown by Moore et al. (1998) and
Diemand et al. (2003), the finite resolution should reduce the halo
concentration. Comparing our results for different boxsizes, it
appears that the resolution effect can lead to an underestimate of
$c$ by $\lesssim 5\%$ at the lower halo mass end in each simulation.
The slight systematic difference between the simulation results and
the ZMJB model predictions may be caused by this resolution effect,
and correcting for this may give a better agreement between the
simulation data and model predictions. We have also examined the
validity of the NFW profile for halos that are in the fast accretion
phase, and found that most of these halos can be fitted by this
profile (Zhao et al. in preparation).

%%%%%%%%%%%%%%%%%%%%%%%%%%%%%%%%%%%%%%%%%%%%%%%%%%%%%%%%%%%%%%%%%%%%%%%
\section{Discussion and conclusions}

We have studied the dependence on mass and redshift of the
concentration of cold dark matter (CDM) halos in high resolution
simulations, and discovered that at early times the mass dependence
of halo concentrations which is pronounced at present, becomes
insignificant, and at $z\ga 3$ halos of mass $> 10^{11}\himsun$ have
in the mean the same density profile with $c\sim 3.5$. Our results
indicate that the median concentration of halos cannot decrease with
redshift or/and halo mass to a value less than $\sim 3$. Massive
cluster halos at present have higher concentrations than some previous
models predicted.

The good agreement between the results of $c$ from different
simulations demonstrates that the simulation resolution effect has
been well controlled in our analysis.

For the mass accretion histories of halos with masses from $10^{10}$
to $10^{15}\himsun$, we have found that they are well expressed by
the universal function [Eq(\ref{mah})].

All these results can be quantitatively matched by the empirical
model of ZMJB. In that model the concentration of a halo is related
to its mass accretion history through the scaling relation they
found for $r_s$ and $M_s$. It predicts that all halos in the fast
accretion phase have similar concentrations, regardless of their
mass or their redshift . The ZMJB model reproduces our results both
in the fast and the slow accretion phases.  Since directly modelling
halo density profiles in numerical simulations is both expensive and
time consuming, the ZMJB model provides a practically useful
technique for modeling internal structures of individual CDM halos.

While the model of Bullock et al. (B01) agrees with our results for
halos in slow accretion, it seriously underestimates the
concentration of halos in the fast accretion phase. The models of
E01 have the same weakness as B01, since they all adopt a similar
assumption that $c(M_{\rm vir},a)\propto a$.

The implications for galaxy formation models, and for the
interpretation of observations, such as strong and weak lensing
surveys, are interesting. According to our results,
massive cluster halos at the present time,
galactic halos at $z\sim 3$, and halos of the first collapsed objects
in the universe at $z\sim 15$ (such as POP III stars) should all have
about the same concentration.

%%%%%%%%%%%%%%%%%%%%%%%%%%%%%%%%%%%%%%%%%%%%%%%%%%%%%%%%%%%%%%%
\acknowledgments
%%%%%%%%%%%%%%%%%%%%%%%%%%%%%%%%%%%%%%%%%%%%%%%%%%%%%%%%%%%%%%%

We are grateful to S. White and L. Gao for helpful discussions.
Numerical simulations presented in this paper were carried out at ADAC
(the Astronomical Data Analysis Center) of the National Astronomical
Observatory, Japan. The work is supported in part by NKBRSF (G19990754)
and by NSFC (No.10125314). We are grateful to the PINOCCHIO team for
making their software publicly available, and to James Bullock and Julio
Navarro for providing their codes to compute $c$.

\end{document}